\documentclass[fleqn,twoside]{article}
\usepackage[headings]{espcrc2}

\readRCS
$Id: espcrc2.tex,v 1.2 2004/02/24 11:22:11 spepping Exp $
\usepackage{graphicx}
\usepackage[figuresright]{rotating}

\newcommand{\AmS}{{\protect\the\textfont2
  A\kern-.1667em\lower.5ex\hbox{M}\kern-.125emS}}

\title{Photoabsorption cross sections at superhigh energies of real photons}

\author{E. V. Bugaev \address{Institute for Nuclear Research,
        Russian Academy of Sciences, \\ 60th October Anniversary
        Prospect 7a, 117312 Moscow, Russia} %
 }

\runtitle{Photoabsorption cross sections at superhigh energies of real photons} %
\runauthor{E.V. Bugaev}

\begin{document}

\begin{abstract}
The brief review of modern theoretical models describing the
process of the photon absorption by nucleons at superhigh energies
of real photons is given. The main aim of the work is an
estimation of the theoretical uncertainty of the cross section
prediction at photon energies around $10^{19}-10^{20} eV$.
\vspace{1pc}
\end{abstract}

\maketitle

\section{Introduction}

Photons of energy $10^{19}-10^{20} eV$ are interesting in
connection with the problem of the origin of high energy cosmic
rays.

The rather abundant fluxes of UHE photons are predicted in
top-down models. UHE photons appear also as a result of
interactions of UHECR with relic radiation.

Search for UHE photons use the comparison of extensive air shower
data with results of detailed simulations based on assumptions on
photonuclear cross section.

Extrapolation of low energy data on $\sigma_{\gamma p}$ has been
given by the parametrization \cite{BezrBug}
\begin{equation}
\sigma_{\gamma p} = 114,3 + 1,647 \; ln^2[s/s_0] \;\;\; (\mu b) ,
\end{equation}
where $s_0 = 88.243 GeV^2$. The $ln^2 s$-law in this formula has
been chosen assuming the validity of the vector dominance model
(VDM) and additive quark model (see below, eq. (\ref{eq9})).

More recently, M. Block and F. Halzen \cite{BlockHal} suggested
the extrapolation of $\sigma_{\gamma p}$ using the analyticity
arguments . They wrote, as a starting point ($\nu$ is the photon
energy in the lab frame):
\begin{equation}
\sigma_{\gamma p} = C_0 + C_1 ln \frac{\nu}{m_p}
 + C_2 ln^2 \frac{\nu}{m_p} + \beta  (\frac{\nu}{m_p})^{\mu-1}.
\end{equation}
Constants $C_i$ are constrained using the precise low energy fit
at $\sqrt{s} < 2,01 GeV$ (given by Damashek and Gilman \cite{DG}).
Authors showed also that the fit with $C_2=0$ (i.e., the $ln\;s$ -
dependence in asymptotic) is not good (from a point of view of the
$\chi^2$-analysis).

\section{Eikonal (minijet) models}


The total photoabsorption cross section in eikonalized minijet
models is calculated by the basic formula (see, e.g., \cite{For})
\begin{eqnarray}
\sigma_{\gamma p} (s) =  2 \frac{\alpha_{em}}{\pi} n_f \langle e^2
\rangle  \int d^2 b \times  \;\;\;\;\;\;\;\;\;\;\;\;\;\;\; \nonumber \\
\;\;\;\;\;\;\;\;\;\;\;\; \times \; \int\limits_{(k_{\perp
0}^{min})^2} \frac {dk_{\perp 0}^2}{k_{\perp 0}^2} [1 -
e^{-\chi(s,b,k_{\perp 0}^2)}].
\end{eqnarray}
Here, $\chi(s,b,k_{\perp 0}^2)$ is the factorised eikonal, in
which the probability that the photon can produce the hadronic
fluctuation ($q\tilde q$-pair) is removed, $k_{\perp 0}$ is the
transverse momentum of the quarks of the pair. In general, the
eikonal function $\chi$ is expressed through the parton densities
inside of the hadronic fluctuation of the photon and inside the
nucleon target,
\begin{eqnarray}
\chi(s,b,k_{\perp 0}^2)=A(b) \int dp_{\perp}^2 \int dx_1 \int dx_2
\times \nonumber
\\  \times \; n_i(x_1, p_{\perp}^2, k_{\perp 0}^2) n_j^p(x_2, p_{\perp}^2)
\frac{d\sigma_{ij}}{dp^2_{\perp}}.
\end{eqnarray}

It is customary to separate the number densities $n_i$ into two
components: a nonperturbative (VMD) component ($k_{\perp 0}<
k_{\perp 0}^0 \sim 1 GeV$) and perturbative one. The latter
corresponds to relatively high masses of the hadronic system
produced by the photon. Further, one can assume that, at least at
not too high energies, the VMD component is dominant and
perturbative component can be neglected. It means that partonic
densities (numbers of small-$x$ gluons) inside vector mesons are
larger than inside of the $q \tilde q$-pair with large $k_{\perp
0}$.

The well-known example of such an approach is the so-called Aspen
model ("QCD-inspired eikonal model"), see \cite{Block} for a
review. In this model the starting point is the eikonal
$\chi(s,b)$ for the case of even scattering hadronic amplitudes
($\frac{1}{2}(f_{pp}+f_{\tilde p p})$) which consists of three
parts:

\begin{equation}
\chi(s,b)= \chi_{qq}(s,b) + \chi_{qg}(s,b) + \chi_{gg}(s,b),
\end{equation}
corresponding to quark-quark, quark-gluon and gluon-gluon
interactions. In particular,
\begin{equation}
\chi_{gg}(s,b) = A(b, \mu_{gg}) \sigma_{gg}(s).
\end{equation}
If $f_g(x) \sim x^{-(1+\epsilon)}$ (gluon structure function), one
has $\sigma_{gg} \sim s^{\epsilon}$.

A factor $A(b, \mu_{gg})$ is the impact parameter distribution.
The parameters $\epsilon$, $\mu_{gg}$, ... are determined from
experiment. The total cross section of proton-proton interaction,
in the unitary (black-body) limit, is
\begin{eqnarray}
\sigma_{tot}(s) \approx 2 \int [1-e^{-\chi_{gg}(s,b)}]d^2 b =
\nonumber
\\ = 2\pi (\frac{\epsilon}{\mu_{gg}})^2 ln^2 \frac{s}{s_0}
\end{eqnarray}
(at asymptotic energies), $A(b,\mu)|_{b\to\infty} \sim e^{-\mu
b}$.

Now, using VMD and additive quark model, the eikonal of $\gamma
p$-scattering can be written as \cite{Block}
\begin{eqnarray}
\chi^{VMD}(s,b) = \frac23\sigma_{gg}(s) A(b, \sqrt{3/2} \;
\mu_{gg}).
\end{eqnarray}
From here one has
\begin{eqnarray} \label{eq9}
\sigma_{\gamma p} = P_{had}^{VMD} \cdot 2 \int [1-e^{-\chi^{VMD}}]
d^2 b \sim ln^2 \frac{s}{s_0},
\end{eqnarray}
where $P_{had}^{VMD}$ is the probability of the photon-vector
meson transition, $P_{had}^{VMD} \sim 4\pi \alpha_{em} /
f^2_{\rho}$.

\section{Regge models}

It is well known that in the Regge approach to deep inelastic
scattering (DIS) the (effective) Pomeron intercept depends on
$Q^2$ and $x_{Bj}$. The $Q^2$-dependence can be connected with DIS
dynamics (e.g., with DGLAP evolution) as well as with the
existence of an additional "hard" Pomeron. Correspondingly, there
are two Regge-type parametrizations of the proton structure
functions, which are frequently used.

In Donnachie-Landshoff model of two Pomerons \cite{DaLo} one has,
in the Regge limit,
\begin{eqnarray} \label{F2}
F_2(x_{Bj}, Q^2)=\sum\limits_i A_i \Big( \frac{Q^2}{Q^2+a_i} \Big)
^ {1+\epsilon_i} x^{-\epsilon_i}_{Bj},
\end{eqnarray}
and, in the photoproduction limit, $\sigma_{\gamma p}$ is finite.
The example of the fit is \cite{DaLo} ($2\nu = s [GeV^2]$):
\begin{eqnarray}
\sigma_{\gamma p} = 0.283(2\nu)^{0.418} + 65.4 (2\nu)^{0.0808}
\;\; [\mu b].
\end{eqnarray}
Authors warn against of using the eq. (\ref{F2}) at extremely
small $x_{Bj}$ where the fixed-power behavior will be moderated by
shadowing suppression.

In the CKMT model \cite{Kaidalov} the main statement is that the
rescattering (absorption) corrections in applications of the Regge
theory to DIS at HERA energies are not small. The Pomeron
intercept $1+\epsilon_0=1.0808$ is not the true Pomeron intercept
(i.e., it does not correspond to the "bare Pomeron"), but rather
is the effective one. The relative contribution of the most
important absorptive corrections depends on $Q^2$. As a result, at
large $Q^2$ we see the bare Pomeron (with intercept $\sim 0.25$)
and at $Q^2=0$ we see the Pomeron with effective intercept
$1.0808$ (i.e., in the photoproduction limit, there is no term
fastly growing with the photon energy).

One should mention also the Regge-type model of \cite{Petru},
where the concept of the effective Pomeron intercept is used, the
value of which is weakly dependent, in the photoproduction limit,
on the photon energy.

\begin{figure}[htb]
\includegraphics[trim=32 40 15 35, width=0.5\textwidth]{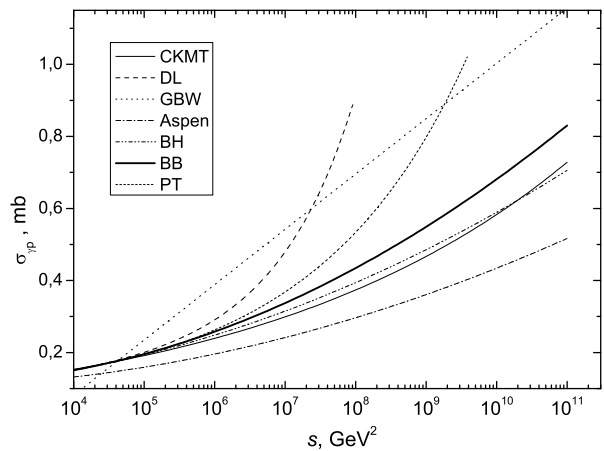}
\caption{The photoabsorption cross section as a function of photon
energy, for different models: CKMT \cite{Kaidalov}, DL
\cite{DaLo}, GBW \cite{GBW}, Aspen \cite{Block}, BH
\cite{BlockHal}, BB \cite{BezrBug}, PT \cite{Petru}.} \label{fig1}
\end{figure}

\section{Colour dipole models}

Real photons and virtual photons with small $x_{Bj}$ have hadronic
properties. The length of hadronic fluctuations is large (and
larger than the target size) at large energies, $l \sim
\frac{1}{m_p x_{Bj}}$ for virtual photons and $l\sim \frac{2
E_{\gamma}}{M^2_{q\tilde q}}$ for real photons ($M_{q\tilde q}$ is
the invariant mass of the hadronic fluctuation). So, the $\gamma
N$-scattering is the two-step process.

The basic equations of the colour dipole model are very simple due
to the fact that colour dipoles are eigenstates of interaction in
QCD and the well known  method of eigenstates \cite{GoodW} can be
used. The total $\gamma p$- cross section is expressed through the
total gluon-nucleon cross section,
\begin{equation}
\sigma_{\gamma p} = \int d^2 r_{\perp} \Psi_{\gamma}(r_{\perp})
\hat \sigma(r_{\perp}).
\end{equation}
Here, $\Psi_{\gamma}(r_{\perp})$ (more exactly, $ \Psi_{T,
L}(r_{\perp}, z, Q^2)$) is the photon-dipole wave function which
depends on the photon virtuality, on the longitudinal momentum
fraction $z$ carried by the quark, and on the dipole size
$r_{\perp}$. These variables are "frozen" during interaction.

The dipole cross section, $\hat \sigma(r_{\perp})$, is assumed to
be flavour-independent and depending, except of $r_{\perp}$, on
$s=W^2$ or $x'$, where $x'$ is the momentum fraction of the proton
carried by the gluon attached to the $q\tilde q$-loop.

Perturbative QCD leads to the formula \cite{Frankfurt}
\begin{equation}
\hat \sigma(r_{\perp}, x') = \frac{\pi^2}{3} r_{\perp}^2
\alpha_s(\bar{Q^2}) x' g(x', \bar{Q^2}),
\end{equation}
where $\bar{Q^2}$ is the energy scale depending on the dipole
size, $g(x', \bar{Q^2})$ is the gluon density.

In MFGS model \cite{MFGS} it is assumed, phenomenologically, that
$ \bar{Q^2} =\frac {\lambda}{r_{\perp}^2} \;\; , \;\; \lambda =
const \;\; (4\div 10). $

Using unitarity, as a guide, the dipole cross section can be
written as
\begin{eqnarray}
\hat \sigma(r_{\perp}, x') = 2 \int d^2 b
[1-e^{-\chi(b,E_{\gamma}, r_{\perp})}] \equiv \nonumber \\
\equiv 2 \int d^2 b \Gamma(E_{\gamma}, r_{\perp}, b).
\end{eqnarray}
The profile factor $\Gamma(E_{\gamma}, r_{\perp}, b)$ is smaller
than $1$. If $\hat \sigma(r_{\perp}, x')$ is known one can
calculate $\Gamma$ assuming some law of $t$-dependence of the
scattering amplitude.

T.Rogers and M.Strikman \cite{RoSt} calculated, using MFGS model,
the $\gamma p$ - cross section up to super-high energies. They
unitarized cross section "by force", calculating profile function
$\Gamma$, and, if the profile function exceeded unity, putting
$\Gamma=1$. By such a way they determined the maximum rise of
$\sigma_{\gamma p}$ with energy, $ \sigma_{\gamma p} \sim ln^3 \;
E_{\gamma}, $ for $E_{\gamma}>10^3 GeV$.

The dipole cross section $\hat \sigma(r_{\perp}, x')$ in MFGS
model rises with energy infinitely. If we suppose that this cross
section is bounded by an energy independent value, as in simple
saturation models \cite{GBW}, the rise of $\sigma_{\gamma p}$ with
energy still will take place, due to the photon wave function. In
the model of \cite{GBW} one assumes that
\begin{eqnarray}
\hat \sigma(x',r_{\perp}) = \sigma_0 [1-e^{-\frac{r_{\perp}^2}{4
R_0^2(x')}}],
\end{eqnarray}
$R_0^2(x') = (x'/x_0)^{\lambda} GeV^{-2}$, $x'\sim x_{Bj}$. If the
Bjorken variable, in the photoproduction limit, is modified to be
\begin{eqnarray}
x=x_{Bj}(1+\frac{4 m_q^2}{Q^2}) \to  %
\frac{4 m_q^2}{W^2} \;\; (Q^2\to 0),
\end{eqnarray}
it is easy to show that $\sigma_{\gamma p} \sim ln  \;
E_{\gamma}$.

The modification of the model \cite{GBW}, with taking into account
the QCD evolution of the gluon distribution \cite{Bartles},
\begin{eqnarray}
\hat \sigma(x',r_{\perp}) = \sigma_0 [1-e^{-\frac{\pi^2
r_{\perp}^2 \alpha_s(\mu^2) x' g(x', \mu^2)}{3 \sigma_0}}]
\end{eqnarray}
(here the scale $\mu^2$ is assumed to have the form $ \mu^2 =
\frac{C}{r_{\perp}^2} + \mu_0^2, $), gives, in asymptotics, $
\sigma_{\gamma p} \sim ln ^ {1/2}  \; E_{\gamma}. $

Physically, the saturation model, in its phenomenological variant,
corresponds to the proton being a disc in the transverse plane
with a sharp bordering. Saturation leads to an uniform blackening
of the disk with decreasing $x_{Bj}$ (for a parent dipole with a
fixed size $r_{\perp}$) without changing the disc size.

For a GBW model \cite{GBW} the parameters are:
\begin{equation}
\lambda=0.3 \;\; ; \;\; x_0=3\times 10^{-4} \;\; ; \;\;
\sigma_0=23 mb.
\end{equation}

\section{Unitary bounds on s-dependence of $\sigma_{\gamma
p}$}

Using three assumptions: {\it i)} $\gamma N$-interaction is a
two-step process, {\it ii)} generalized VDM (a dispersion relation
with variable $M^2$, where $M$ is the mass of the hadronic
fluctuation), {\it iii)} the hadronic interaction of the
fluctuation ($q\tilde q$-pair) is a black disc interaction, one
can show that the $\sigma_{\gamma^* N}$ is given by \cite{Gribov}
\begin{equation}
\sigma_{\gamma^* N} = \frac{\alpha_{em}}{3\pi}\int\frac{R(M^2)M^2
d M^2}{(Q^2+M^2)^2} \sigma_{M^2 N}(s),
\end{equation}
where $R = \frac{\sigma(e^+ e^- \to hadrons)} {\sigma(e^+ e^- \to
\mu^+ \mu^-)}.$ If we suppose that the hadronic cross section
$\sigma_{M^2 N}$ is given by the black-body limit, $ \sigma_{M^2
N}  \sim ln ^ {2} \frac {s} {s_0}$, one obtains, in the
photoproduction limit,
\begin{equation}
\sigma_{\gamma N} \sim ln^2 \frac{s}{s_0} \; ln
\frac{M_{max}+Q^2}{M_{min}+Q^2} \Big|_{Q^2\to 0} \sim ln^3
\frac{s}{s_0}.
\end{equation}
It is the so called Gribov bound \cite{Abr}. However, the
hypothesis of the black disk interaction cannot be correct for
$q\tilde q$-pairs with large mass $M_{q\tilde q}$. Such a pair has
a small $r_{\perp} \sim 1/M_{q\tilde q}$ and, being colour
neutral, interacts with the target weakly, $\sigma \sim
r_{\perp}^2$ (it is predicted by pQCD). Due to this, there is the
following constraint on the value of $M$ \cite{Gotsman}:
\begin{equation}
M^2 |_{max} \sim \Lambda^2  e^{\sqrt{a ln(1/x_{Bj})}}
\end{equation}
($\Lambda$ is the QCD scale), so, in far asymptotics one obtains
the corrected unitary bound (assuming that $x_{Bj}^{min} \sim
m_q^2 / s$, as in eq. (16)) \cite{Gotsman}:
\begin{equation}
\sigma_{\gamma N} \sim ln^2 \Big(\frac{s}{s_0}\Big)
ln^{1/2}\Big(\frac{1}{x_{Bj}^{min}}\Big) \sim ln ^{2.5}
\frac{s}{s_0}.
\end{equation}



\section{Conclusions}

The straggling of theoretical predictions for $\sigma_{\gamma p}$
at $E_{\gamma} \sim 10^{19}-10^{20} eV$ is large, but not
catastrophically. At $s=10^{11} GeV^2$ the predictions are in
interval $ (0.5 \div 1.1) mb. $

It is rather difficult to predict reliably the asymptotic
$s$-dependence of the $\sigma_{\gamma p}$. Most probably, the
upper limit on the rise of the photoabsorption cross section is
given by the law $\sim ln^3 E_{\gamma}$ as follows from the
estimates of authors of \cite{RoSt} based on the colour dipole
model. The corrected Gribov unitary bound based on generalized VDM
and pQCD constarint gives $\sigma_{\gamma p} \sim ln^{2.5}
E_{\gamma}$. Predictions based on the vector meson dominance and
additive quark model give the law $\sim ln^2 E_{\gamma}$. At last,
Regge eikonal model \cite{Kaidalov} predicts, asymptotically,
$\sigma_{\gamma p} \sim E_{\gamma}^{0.1}$.

\end{document}